\documentclass[aps,preprint,showpacs,floatfix]{revtex4}
\usepackage{graphicx,bm,amsmath,dcolumn,}
\usepackage{verbatim}
\begin{document}
\newcommand{\be}{\begin{equation}}
\newcommand{\ee}{\end{equation}}
\newcommand{\half}{\frac{1}{2}}
\newcommand{\ith}{^{(i)}}
\newcommand{\im}{^{(i-1)}}
\newcommand{\gae}
{\,\hbox{\lower0.5ex\hbox{$\sim$}\llap{\raise0.5ex\hbox{$>$}}}\,}
\newcommand{\lae}
{\,\hbox{\lower0.5ex\hbox{$\sim$}\llap{\raise0.5ex\hbox{$<$}}}\,}

\title{Berezinskii-Kosterlitz-Thouless-like percolation 
transitions in the two-dimensional XY model}

\author{ Hao Hu$^1$, Youjin Deng$^1$~\footnote{Email: yjdeng@ustc.edu.cn}, 
and Henk W. J. Bl\"ote$^{2,3}$ } 
\affiliation{$^{1}$ Hefei National Laboratory for Physical
Sciences at Microscale,
Department of Modern Physics, University of Science and
Technology of China, Hefei 230027, China }
\affiliation{$^{2}$ Instituut Lorentz, Leiden University,
P.O. Box 9506, 2300 RA Leiden, The Netherlands}
\affiliation{$^{3}$Faculty of Applied Sciences, Delft University of
Technology, P. O. Box 5046, 2600 GA Delft, The Netherlands}
\date{\today} 

\begin{abstract}
We study a percolation problem on a substrate formed by two-dimensional
XY spin configurations, using Monte Carlo methods. 
For a given spin configuration we construct percolation clusters by
randomly choosing a direction $x$ in the spin vector space, and then
placing a percolation bond between nearest-neighbor sites $i$ and $j$ 
with probability $p_{ij}=\max (0,1-e^{-2K s^x_i s^x_j})$, where $K > 0$ 
governs the percolation process.  A line of percolation thresholds
$K_{\rm c} (J)$ is found in the low-temperature range $J \geq J_{\rm c}$, 
where $J > 0$ is the XY coupling strength.
Analysis of the correlation function $g_p (r)$, defined as the
probability that two sites separated by a distance $r$ belong to the
same percolation cluster, yields algebraic decay for $K\geq K_{\rm c}(J)$,
and the associated critical exponent depends on $J$ and $K$.
Along the threshold line $K_{\rm c}(J)$, the scaling dimension for $g_p$
is, within numerical uncertainties, equal to $1/8$.
On this basis, we conjecture that the percolation transition along
the $K_{\rm c} (J)$ line is of the Berezinskii-Kosterlitz-Thouless type.
\end{abstract}

\pacs{05.50.+q(lattice theory and statistics), 64.60.ah(percolation),
 64.60.F-(equilibrium properties near critical points, critical exponents),
 75.10.Hk(classical spin models)}

\maketitle

\section{Introduction}
The XY model is formulated in terms of two-dimensional spins $\vec{s}$
normalized as $|\vec{s}|=1$, residing on the sites of a lattice.
The reduced Hamiltonian of the XY model (already
divided by $k_{\rm B} T$ with $k_{\rm B}$ the Boltzmann constant and $T$ the
temperature) reads 
\begin{equation}
\mathcal{H} \;=\; -J \sum\limits_{\langle i j \rangle } 
\vec{s}_i \cdot \vec{s}_j \; ,
\label{def_H}
\end{equation}
where the sum is over all nearest-neighbor pairs, and $J > 0$ is the
ferromagnetic coupling strength. The spins are labeled by their site
numbers.

It is known from the Mermin-Wagner-Hohenberg-Coleman
theorem~\cite{Mermin-66} that there cannot exist spontaneous long-range 
order as long as $J$ is finite in Eq.~(\ref{def_H}), because thermal
fluctuations are strong enough to destroy the order.  Nevertheless,
the system undergoes a phase transition \cite{B,KT,Kosterlitz-74} as
the coupling strength $J$ increases.
This type of transition is of infinite order and is known as the
Berezinskii-Kosterlitz-Thouless (BKT) transition.  For $J < J_{\rm c}$,
the spin-spin correlation function decays exponentially, and the spins
form a plasma of vortices; but for $J > J_{\rm c}$, the spin-spin
correlation function decays algebraically with an exponent depending on
$J$, and the spin configurations contain bound vortex-antivortex pairs.
Transitions of the BKT type occur in various kinds of systems. The
XY-type of transition is related by duality to roughening transitions
in solid-on-solid and related models \cite{JK}. Apart from the XY model,
BKT transitions are found, among others, in vertex models \cite{6v},
models of crystal surfaces~\cite{vB}, the antiferromagnetic triangular
Ising model~\cite{NHB}, string theory~\cite{Maggiore-02},
network systems~\cite{network}, superfluid systems~\cite{Bishop-78},
and superconducting systems~\cite{Resnick-81}.
These models may involve long-range or short-range interactions. 

It is also known that certain observables of statistical models  are
equivalent or closely related to properly defined geometric quantities. 
For instance, the Potts model can be exactly mapped onto the random-cluster
model~\cite{KF}, and the susceptibility $\chi$ of the former is related to
the cluster-size distribution of the latter; a similar situation applies
to the Nienhuis O($n$) loop model \cite{N82} and the equivalent spin 
model~\cite{Domany-81}. The Mott-to-superfluid transition in the
Bose-Hubbard model can be characterized by the winding number of the
world lines of the particles~\cite{Pollock-87}.
The geometric percolation~\cite{Stauffer-94} process has been employed
to study percolation on critical substrates, such as the Ising
model~\cite{Sykes-76,Coniglio-77,Deng-04}, the Potts model~\cite{Qian-05},
the O($n$) model~\cite{Ding-09} and even quantum Hall systems~\cite{Lee-93}.
 
In this work, we study the percolation problem on the substrate of the
XY model~(\ref{def_H}). There is still some freedom in the choice of the
percolation criterion. For instance, one may place percolation bonds
between all neighboring XY spins if their orientations differ less than
a given angle called the ``conducting angle''. This problem was recently
investigated by Wang et al.~\cite{Wang-10}.
Here we use a different criterion. For a given spin configuration,
we choose a randomly oriented Cartesian reference frame $(x,y)$ in the
two-dimensional spin space, and place bonds between nearest-neighbor
pairs, say sites $i$ and $j$, with a probability
\begin{equation}
p_{ij}=\max (0, 1-e^{-2K s^x_i s^x_j}) \, ,
\label{pij}
\end{equation}
where $K > 0$ parametrizes
the percolation problem. Note that for $K=J$ these percolation clusters 
reduce to those formed by the cluster simulation process of the XY
model as described in Sec.~\ref{sec_alg}.

The rest of this work is organized as follows.  Section~\ref{sec_cri}
presents our numerical results for the critical points of the XY model
on the square as well as on the triangular lattice.
Section~\ref{sec_per} describes the analysis of the percolation problem
for both lattices, with an emphasis on the determination of the
universal character of this type of percolation transition.
We conclude with a discussion in Sec.~\ref{sec_dis}. 

\section{Algorithm and Sampled Quantities}\label{sec_alg}
\subsection{Spin updating algorithm}
We employ efficient Monte Carlo simulations of the XY model~(\ref{def_H})
by means of a cluster method \cite{Swendsen-87,Wolff}. We use a full
cluster decomposition~\cite{Swendsen-87} as follows:
\begin{enumerate}
\item Choose a randomly oriented Cartesian frame of reference ($x,y$)
in the {\em spin} space, and project the spin along the $x$ and $y$
axes as $\vec{s} =s^x \hat{x} +s^y \hat{y}$.
Accordingly, the scalar product in Eq.~(\ref{def_H}) is written
$\vec{s}_i \cdot \vec{s}_j=s^x_i s^x_j + s^y_i s^y_j$, so that 
the Hamiltonian separates into two parts as $\mathcal{H} = 
\mathcal{H}_x + \mathcal{H}_y$, with $\mathcal{H}_x=-J 
\sum\limits_{\langle i j \rangle } s^x_i s^x_j$ and similar 
for $\mathcal{H}_y$.
\item Between each pair of nearest-neighboring sites, e.g., site $i$ and $j$,
place a bond with probability $p_{ij} =\max(0, 1-e^{-2J s^x_i s^x_j})$.
\item Construct clusters on the basis of the occupied bonds.
\item Independently for each cluster, flip the $x$ components of 
all the spins in the cluster with probability $1/2$.
\end{enumerate}

\subsection{Sampled quantities}
We sampled several quantities, including
the second and the fourth moments of the magnetization density, 
${M}^2=|\sum_k \vec{s}_k|^2 /V^2$ and ${M}^4=|\sum_k \vec{s}_k|^4 /V^4$.
These quantities determine the dimensionless Binder
ratio~\cite{Binder-81} as 
\begin{equation}
Q_m = \langle {M}^2 \rangle^2 / \langle {M}^4 \rangle \; ,
\label{def_Qm}
\end{equation}
where $V=L^2$ is the volume.
The susceptibility  is $\chi=V \langle {M}^2 \rangle $.

Denoting the size of the $i$th cluster by ${C}_i$, we also sampled the
second and the fourth moments of the cluster size distribution as 
\begin{equation}
{S}_2 = \frac{1}{V^2} \sum_{i} {C}_i ^2 \hspace{5mm} 
\mbox{and} \hspace{5mm}
{S}_4 = \frac{1}{V^4} \sum_{i} {C}_i ^4 \; .
\label{def_S2}
\end{equation}
Accordingly, we define another dimensionless ratio as 
\begin{equation}
Q_{l} =\langle {S}_2 \rangle^2/ (3 \langle 
{S}_2^2 \rangle- 2 \langle {S}_4 \rangle)  \; .
\label{def_Ql}
\end{equation}
Note that for the Ising model $Q_{l}$ in Eq.~(\ref{def_Ql}) is equal
to $Q_{m}$ in Eq.~(\ref{def_Qm}). 

Also the spin-spin correlation $g_s (r)$ over distances $r=L/2$ and $L/4$
was sampled. A third dimensionless ratio $Q_s$ is defined as 
\begin{equation}
Q_s= \langle g_s (L/2) \rangle /  \langle g_s (L/4) \rangle \, .
\label{def_Qs}
\end{equation}
In the high-temperature range $J < J_{\rm c}$, the spin-spin correlation
decays exponentially, and $Q_s$ goes to 0 as $L \rightarrow \infty$. 
At criticality, however, $g_s(r)$ tends to algebraic decay as $ r^{-2x_h}$,
and $Q_s $ converges to  a nontrivial universal value. In an ordered state
with a non-zero magnetization density, $Q_s$ would converge to 1 instead.

Finally, we define the correlation function $g_p(r)$ as the probability
that two sites at a distance $r$ belong to the same cluster. We sampled
$g_p$ over distances $r=L/2$ and $r=L/4$.
The associated dimensionless ratio is defined as 
\begin{equation}
Q_p= \langle g_p (L/2) \rangle /  \langle g_p (L/4) \rangle \, .
\label{def_Qp}
\end{equation}

\section{Critical points}\label{sec_cri}
\subsection{Square lattice}
\begin{figure}[ht]
\vspace*{0cm} \hspace*{-0cm}
\begin{center}
\includegraphics[width=8.5cm]{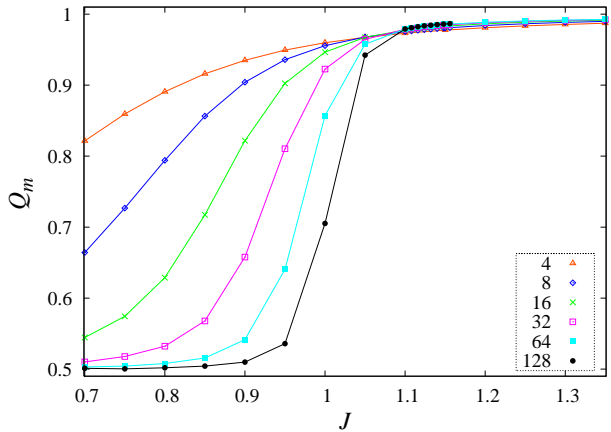}
\caption{(color online). Binder ratio $Q_m$ vs.
coupling strength $J$ for the square lattice. The lines connecting data
points are added for clarity.}
\label{fig_qmsq}
\end{center}
\end{figure}

We simulated the XY model on $L \times L$ square lattices with periodic
boundary conditions, with system sizes in the range $4\leq L\leq 1024$. 
As usual in Monte Carlo studies, the location of a critical point can
well be determined using a dimensionless ratio. This is shown in
Fig.~\ref{fig_qmsq} for the Binder ratio $Q_m$. 
For $J < J_{\rm c}$, $Q_m$ approaches the infinite-temperature value 
$1/2$ as $L \rightarrow \infty$, as expected for a normal distribution
of the $x$ and $y$ components of the magnetization. For $J > J_{\rm c}$,
$Q_m$ rapidly converges to a temperature-dependent value,
as expected in the low-temperature XY phase.

\begin{figure}[ht]
\vspace*{0cm} \hspace*{-0cm}
\begin{center}
\includegraphics[width=7.5cm]{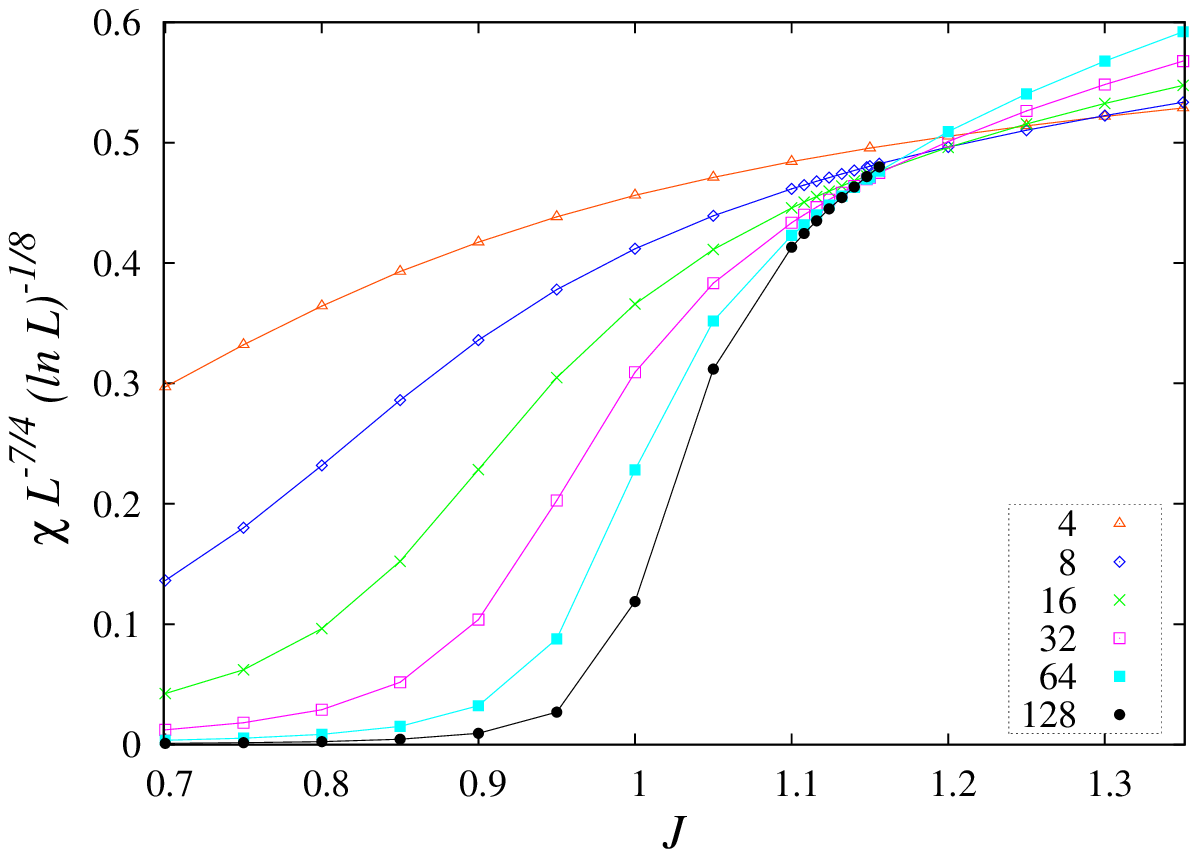}\\
\includegraphics[width=7.5cm]{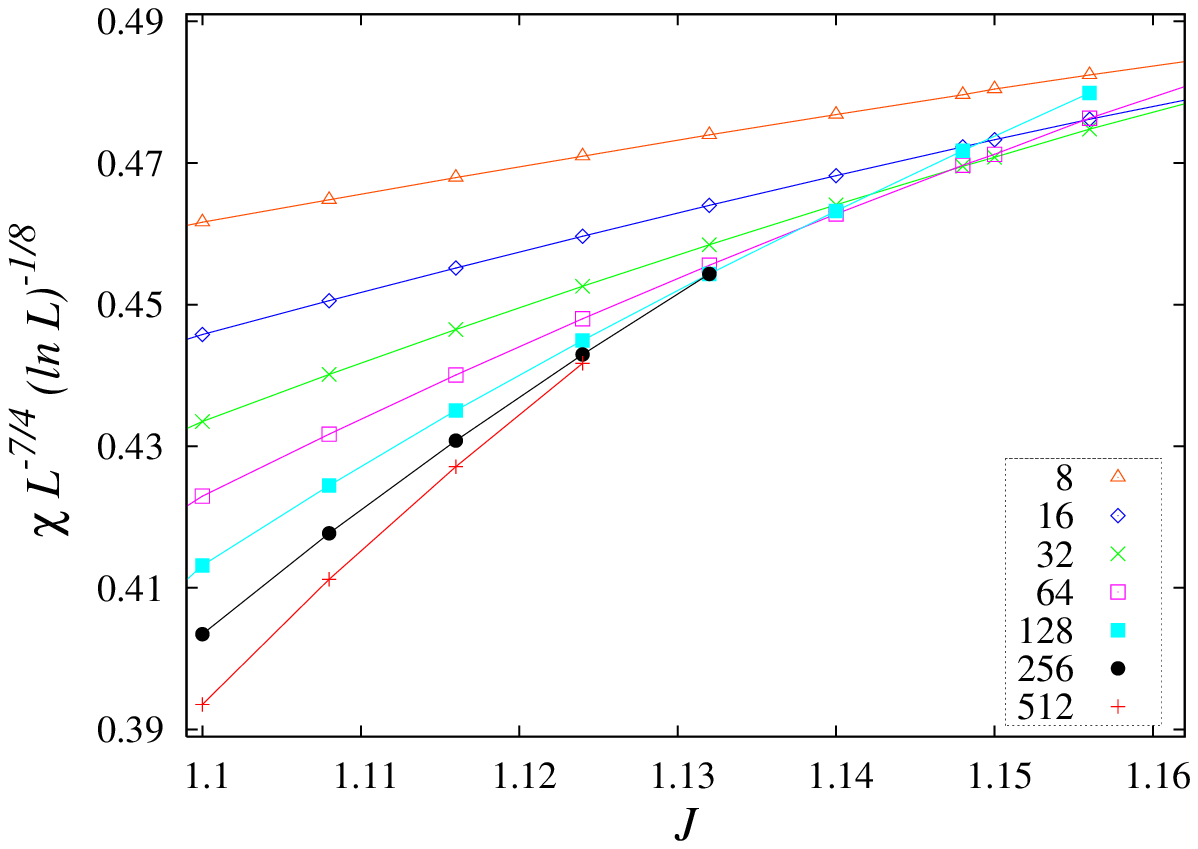}
\caption{(color online). 
Scaled susceptibility $\chi L^{2x_h-2} (\ln L)^{-1/8}$ vs. 
coupling strength $J$ for the square lattice, with $x_h=1/8$. The lower
figure is an enlarged version, and includes data for $L=256$ and $512$.
The lines connecting the data points are added for clarity.  } 
\label{fig_m2sq}
\end{center}
\end{figure}

Making use of the known magnetic scaling dimension 
$x_h=1/8$~\cite{Kosterlitz-74} at the BKT transition, and the logarithmic
correction factor with exponent $1/8$~\cite{Kosterlitz-74,logarithmic}, 
we expect that the scaled quantity $\chi L^{2x_h-2} (\ln L)^{-1/8}$ tends
to a constant at the transition point. The intersections in
Fig.~\ref{fig_m2sq}, which shows this scaled quantity as a function of
$J$ for several system sizes, confirm this expectation.

Using the least-squares criterion, we fitted the quantity
$\chi L^{2x_h-2} (\ln L)^{-1/8}$ data by the formula
\begin{eqnarray}
\chi L^{2x_h-2} (\ln L)^{-1/8} &=& a_0 + \sum_{i=1}^{3} a_i (J_{\rm c}-J)^i
(\ln L)^i + \sum_{j=1}^{2} r_j (J_{\rm c}-J)^j \nonumber \\
&+& b_1/(\ln L) + b_2 L^{-1} +b_3 L^{-2} \; ,
\label{fit_m2}
\end{eqnarray}
where the multiplicative and additive logarithmic corrections have been 
taken into account.  We find that
the data for $ 16 \leq L \leq 1024$ and $ 1.100 \le J \leq 1.125$ are
well described by Eq.~(\ref{fit_m2}).
The fit yields $J_{\rm c}=1.124~(3)$. 

\begin{figure}[htp]
\vspace*{0cm} \hspace*{-0cm}
\begin{center}
\includegraphics[width=8.5cm]{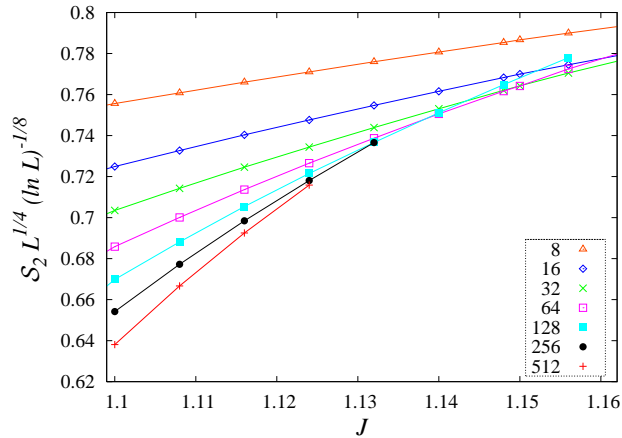}
\caption{(color online).
Scaled second moment ${S}_2 L^{2x_h} (\ln L)^{-1/8}$ of the cluster
size distribution versus coupling strength $J$. These results apply
to the cluster decomposition of the spin model on the square lattice.
The lines connecting the data points are added for clarity.}
\label{fig_s2sq}
\end{center}
\end{figure}

For the Ising model, one can prove that
$\chi=V \langle {M}^2 \rangle = V \langle {S}_2 \rangle$,
which exactly relates the thermodynamic quantity $\chi$ to the geometric
quantity ${S}_2$.  We thus expect that, in the case of the XY model,
the singularity of $S_2$ coincides with that of $\chi$.
The data for ${S}_2 L^{2x_h} (\ln L)^{-1/8}$ (shown in Fig.~\ref{fig_s2sq})
were fitted by Eq.~(\ref{fit_m2}). This fit yields $J_{\rm c}=1.120~(9)$, 
consistent with the result from $\chi$.

There exist already many estimates for the critical point of the
XY model on the square lattice, the latest of which are $J_{\rm c}=1.1199(1)$
by Martin and Klaus~\cite{Martin-05}, $J_{\rm c}=1.1198~(14)$ 
by Butera and Pernici~\cite{Butera-08},
and $J_{\rm c}=1.1200~(1)$ by Arisue~\cite{Arisue-09}.
Our result for $J_{\rm c}$ is consistent with these existing values.

\subsection{Triangular lattice}
\begin{figure}[t]
\vspace*{0cm} \hspace*{-0cm}
\begin{center}
\includegraphics[width=8.5cm]{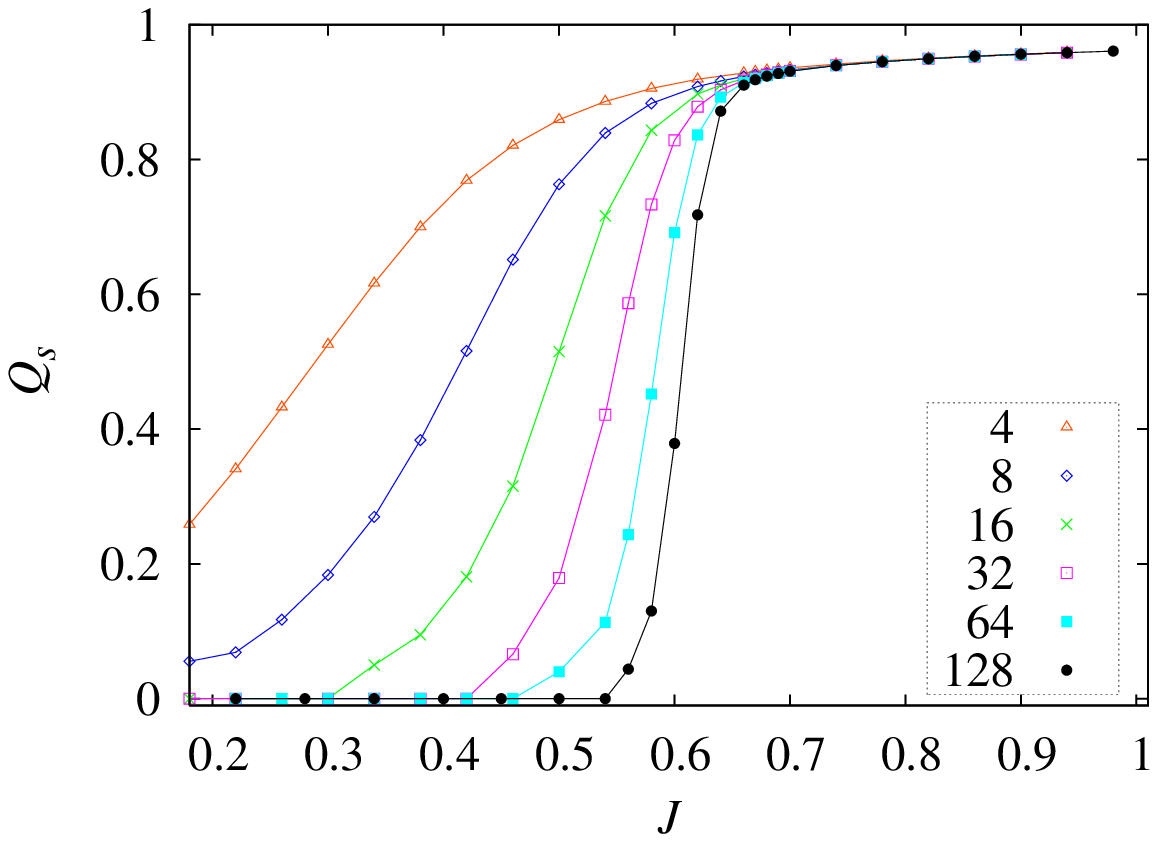}
\caption{(color online).
Ratio $Q_s=g_s ( r=L/2) /g_s ( r=L/4) $ 
vs. coupling strength $J$ for the triangular lattice, with $g_s$ the 
spin-spin correlation. The lines connecting data
points are added for clarity.}
\label{fig_qstr}

\includegraphics[width=8.5cm]{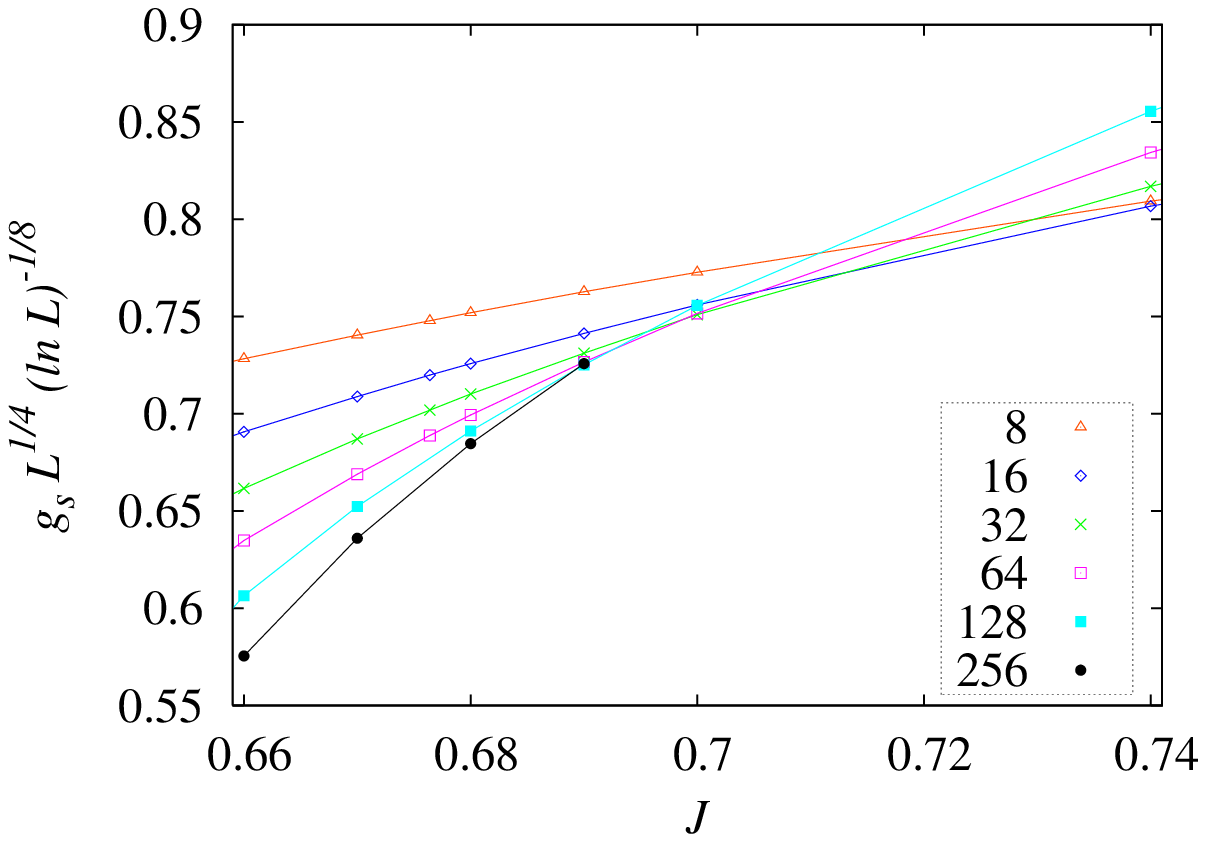}
\caption{(color online).
Scaled correlation $g_s ( r=L/2) L^{2x_h} (\ln L)^{-1/8} $
 vs. coupling strength $J$ for the triangular lattice, with $x_h=1/8$ . 
The lines connecting the data points are added for clarity.}
\label{fig_gstr}
\end{center}
\end{figure}

We also simulated the XY model on the triangular lattice with periodic
boundary conditions, for linear system sizes $L$ in the range 
$4 \leq L \leq 512$. The BKT phase transition is clearly exposed by
Fig.~\ref{fig_qstr}, which plots the ratio $Q_s=g_s(r=L/2)/g_s(r=L/4)$
versus the coupling strength $J$. In the high-temperature range
$J < J_{\rm c}$, $Q_s$ rapidly approaches zero, which reflects the
absence of long-range correlations; in the low-temperature range
$J > J_{\rm c}$, it converges to a $J$-dependent value smaller than 1,
in agreement with the presence of algebraically decaying correlations 
and the absence of a spontaneous magnetization.

The $g_s (L/2) L^{2x_h} (\ln L)^{-1/8}$ data near criticality are shown 
in Fig.~\ref{fig_gstr}. They were fitted by Eq.~(\ref{fit_m2}), which
yielded $J_{\rm c}=0.6833~(6)$. Analogous analyses were performed
 for the scaled susceptibility $\chi L^{2 x_h-2} (\ln L)^{-1/8}$, 
leading to $J_{\rm c}=0.6831~(6)$. Our results for
the critical coupling are consistent with the latest result
$J_{\rm c}=0.6824~(8)$ by Butera and Pernici~\cite{Butera-08}.

\section{Percolation analysis}\label{sec_per}

For each spin configuration generated by the Monte Carlo algorithm,
we performed a full decomposition in percolation clusters, using the
randomly oriented Cartesian frame in the spin space as chosen in the
preceding Monte Carlo step, and then placing 
bonds between nearest-neighbor pairs with probabilities
$p_{ij} =\max(0, 1-e^{-2K s^x_i s^x_j})$. The variable parameter $K > 0$
governs the percolation process. While these percolation clusters are not
involved in spin-updating, they reduce to those obtained during the
cluster simulations in the case $K=J$. To analyze this percolation
problem, we sampled several quantities, including the second and fourth
moments ${S}_2$ and ${S}_4$ of the cluster size distribution, the Binder
ratio $Q_l$, the correlations $g_p (r=L/4)$, $g_p (r=L/2)$, and the ratio
$Q_p$. 
In this Section we describe the numerical results and analyses, 
and also an exact result for the present percolation problem
on the triangular lattice. 

\subsection{Percolation on the square lattice}\label{persq}
\subsubsection{High-temperature range}\label{persq_htr}

For $\tanh K = 1$, i.e. in the limit $K \rightarrow \infty$, all pairs
of nearest-neighbor spins are connected as long as their $x$ components 
are pointing in the same direction.
At zero coupling $J = 0$, spins at different sites are uncorrelated,
so that the percolation process reduces to standard site-percolation
process, since the  site occupation probability $p=1/2$ may be 
identified with the sign of $s^{x}$.
An unimportant difference is that the present process forms
percolation clusters for all the lattice sites while the 
standard site percolation constructs clusters only for the occupied sites.
The site-percolation threshold $p_{\rm c}^{\rm s}$ on the square lattice is 
very close to 0.592746 \cite{Newman-00,ML,FDB}, and thus no infinite
percolation cluster can occur at zero coupling strength $J = 0$, even
for $\tanh K=1$.  Furthermore, from the results in Ref.~\cite{Qian-05},
where a similar percolation problem is studied in the context of 
several Potts models, we expect that no percolation transition
occurs on the square lattice for small $J$.
This expectation was confirmed by Monte Carlo simulations that were
performed at several nonzero $J < J_{\rm c}$. Variation of $K$ did not yield
any signs of a percolation threshold.

\subsubsection{Low-temperature range}\label{persq_ltr}

The low-temperature XY phase $J \geq J_{\rm c}$ displays algebraically
decaying spin-spin correlations, which, unlike the exponential decay at
$J < J_{\rm c}$, allows the formation of a divergent percolation cluster
for sufficiently large $K$. We may thus expect a percolation threshold
to occur at a $J$-dependent value $K_{\rm c} (J)$. 

\begin{figure}[t]
\vspace*{0cm} \hspace*{-0cm}
\begin{center}
\includegraphics[width=7.5cm]{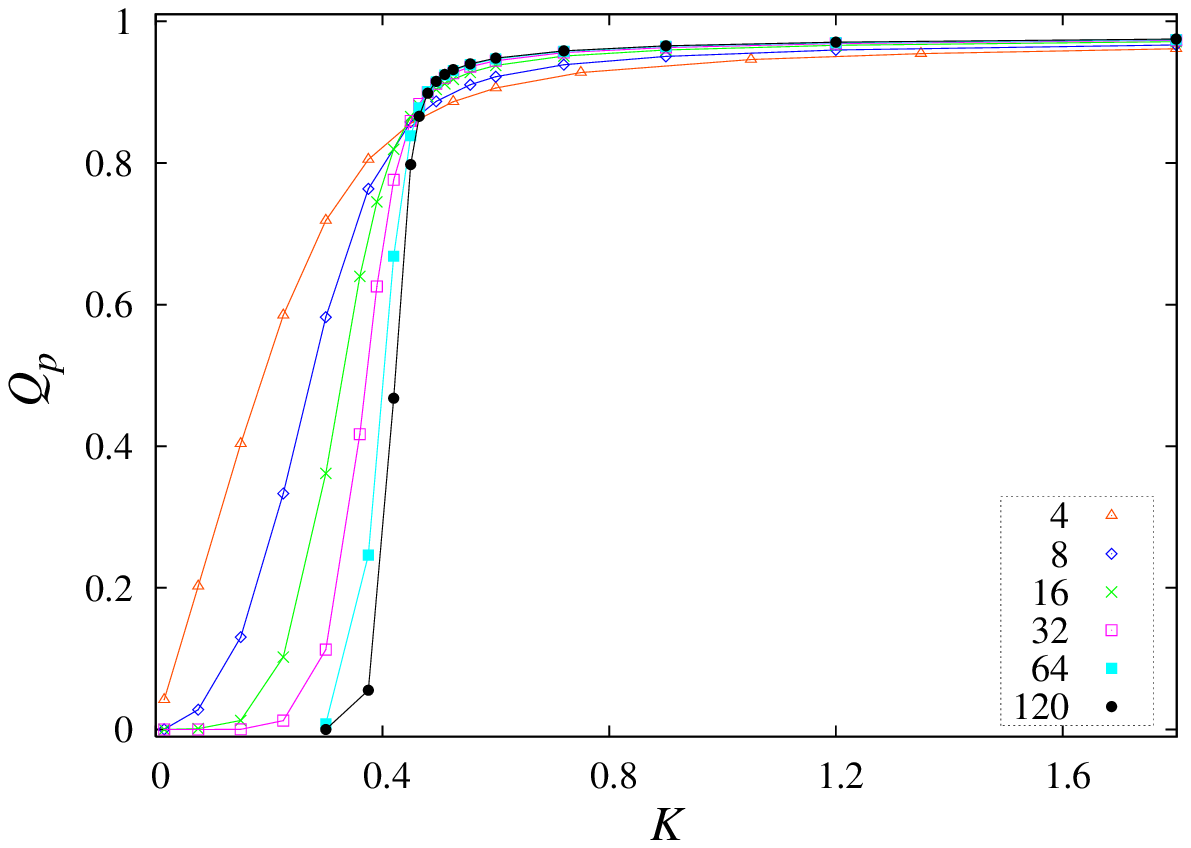}\\
\includegraphics[width=7.5cm]{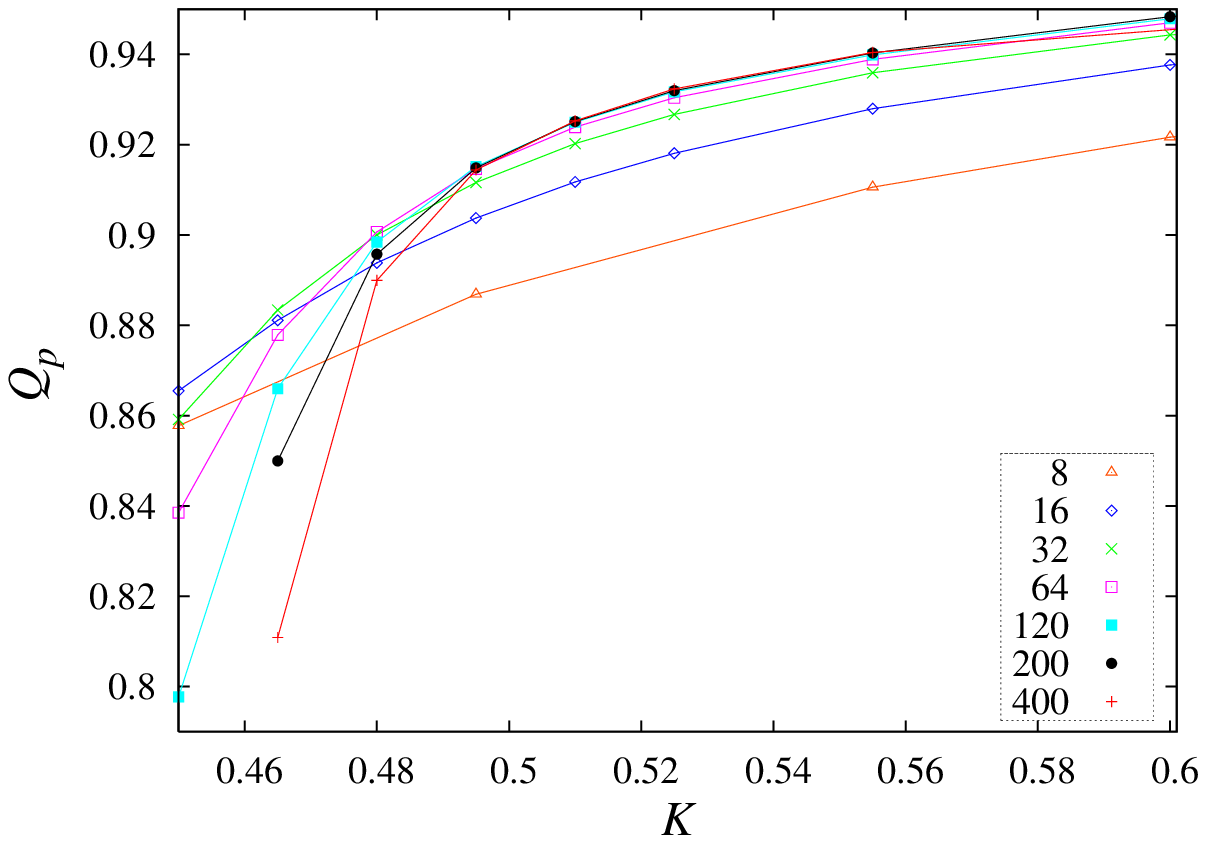}

\caption{(color online).
Dimensionless ratio $Q_p$  vs. the parameter $K$ inducing the percolation
transition. These data apply to the model on the square lattice, with
spin coupling $J=3.0$. The lower figure is an enlarged version.
The lines connecting the data points are added for clarity.}
\label{fig_qpsq}
\end{center}
\end{figure}

The existence of a percolation threshold $K_{\rm c}(J)$ for $J>J_{\rm c}$
is shown by the intersections of the curves in Fig.~\ref{fig_qpsq},
which displays $Q_p$ as a function of $K$ at $J=3.0$ for several $L$.
These data show that $ K_{\rm c} (J=3.0) \approx 0.505 $.
For $K < K_{\rm c} (J) $, $Q_p$ rapidly approaches zero, as expected
from the absence of long-range correlations of $g_p (r)$. 

In view of the long-range spin-spin correlations 
for $J \geq J_{\rm c}$, we have no reason to expect that the percolation
transition at the threshold $K_{\rm c} (J)$ for $J > J_{\rm c}$ belongs to
the uncorrelated percolation universality class. This is supported by the 
observation that, at $J=J_{\rm c}$, the fractal dimension of clusters with
$K=J_{\rm c}$ is $2-x_h=15/8$, which is different from the value 91/48 for
critical percolation clusters~\cite{Domb-87}.
A closer look at the plot of $Q_p$ vs.~$K$ (Fig.~\ref{fig_qpsq}) 
indicates that, for $K \geq  K_{\rm c} (J)$, $Q_p$ rapidly converges to a
$K$-dependent nontrivial value smaller than 1.
We propose the interpretation that, like the thermal transition
induced by the variation of $J$, the percolation transition
induced by $K$ is also BKT-like. 

\begin{figure}[t]
\vspace*{0cm} \hspace*{-0cm}
\begin{center}
\includegraphics[width=7.5cm]{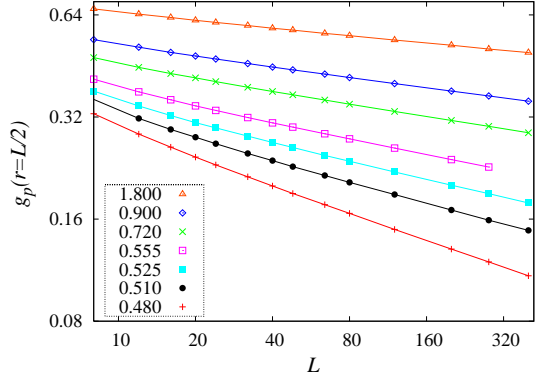}
\caption{(color online).
Correlation $g_p (L/2)$ vs. linear system size $L$ for several values of
$K$ which are shown in the inset. The use of logarithmic scales displays 
the approximate power-law dependence on $L$. These data apply to the 
square-lattice XY model at $J = 3.0$. The lines connecting the data points
are added for clarity.}
\label{fig_gpsq1}
\end{center}
\end{figure}

In Fig.~\ref{fig_gpsq1} we display the correlation $g_p$ over a
distance $r=L/2$ as a function of the linear system size $L$ for
several values of $K$. This figure shows a dependence of $g_p$ on
$L$ that approaches power-law behavior for large $L$.
This suggests that percolation clusters remain  critical for
$K>K_{\rm c}$. Furthermore the 
exponent governing the scaling of $g_p (r)$ appears to depend on $K$.

\begin{figure}[t]
\vspace*{0cm} \hspace*{-0cm}
\begin{center}
\includegraphics[width=7.5cm]{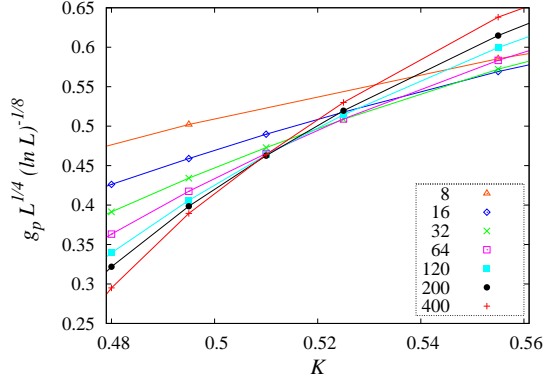}
\caption{(color online).
Scaled correlation $g_p L^{1/4} (\ln L)^{-1/8} $ over a distance
$r=L/2$ vs. $K$ for various linear system size $L$ shown in the inset.
These data apply to the square-lattice XY model at $J=3.0$. 
The lines connecting the data points are added for clarity.}
\label{fig_gpsq2}
\end{center}
\end{figure}

\begin{figure}[t]
\vspace*{0cm} \hspace*{-0cm}
\begin{center}
\includegraphics[width=7.5cm]{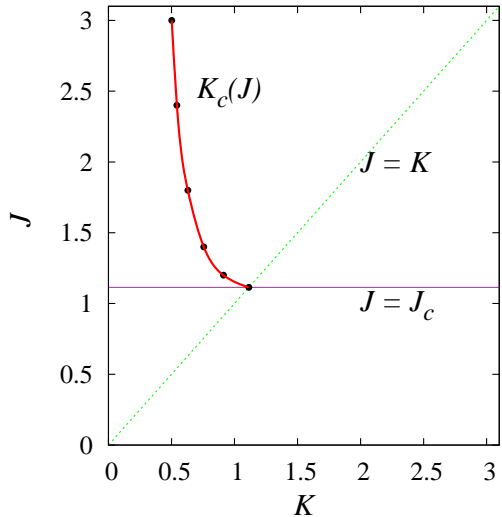}
\caption{(color online).
Phase diagram of the square-lattice XY model in the $J-K$ parameter
space.  The horizontal line represents the thermal BKT transition,
and the diagonal line applies to $K=J$, where the percolation
clusters are just those formed by the cluster algorithm.
The line connecting the data points is added for clarity.}
\label{fig_phase}
\end{center}
\end{figure}

\begin{figure}[t]
\vspace*{0cm} \hspace*{-0cm}
\begin{center}
\includegraphics[width=8.0cm]{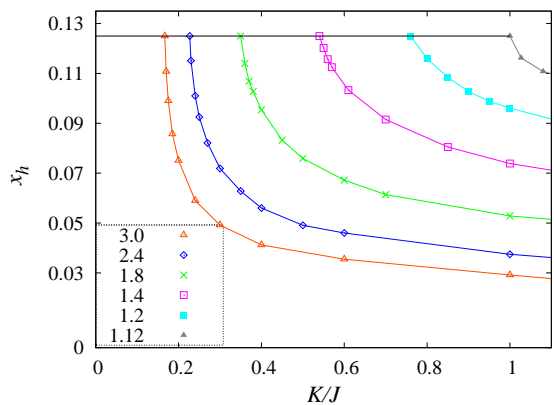}
\caption{(color online).
Scaling dimension $x_h$ for percolation clusters at various 
points $(J,K)$ for the square lattice. The values of $J$ are shown in 
the inset. The horizontal line is located at $x_h=1/8$ and corresponds
with the critical line $K_c(J)$.
This figure indicates that the scaling dimension $x_h$ depends on $K$
as well as on $J$, and approaches $1/8$ for $K \to K_{\rm c}(J)$.
The lines connecting the data points are added for clarity.}
\label{fig_xh}
\end{center}
\end{figure}

\begin{table}[htp]
\vspace*{0cm} \hspace*{-0cm}
\begin{center}
\caption{Results for the scaling dimension $x_h$ at $J=3.0$ for 
various values of $K$ for the square lattice.  Parameters $L_{\rm min}$ 
and $L_{\rm max}$ are the minimum and the maximum system
size between which the Monte Carlo data of $g_p$ are included in the fit.}
\label{tab_xJ3}
\begin{tabular}{c|ccccc}
\hline\hline
$K$   & $3.0$        &$1.8$        &$1.2$   &$0.9$            &$0.72$      \\
\hline    
$L_{\rm min}$ &$20$  &$20$         &$16$         &$16$        &$12$        \\
$L_{\rm max}$ &$400$ &$400$        &$400$        &$400$       &$400$       \\
$x_h$&$0.02916~(2)$  &$0.0345~(6)$ &$0.0412~(6)$ &$0.0492~(6)$&$0.0590~(6)$\\
\hline \hline 
$K$   &$0.60$        &$0.555$      &$0.525$  &$0.510$ &$0.495$             \\
\hline 
$L_{\rm min}$ &$16$  &$16$         &$16$         &$16$        &$16$        \\
$L_{\rm max}$ &$400$ &$400$        &$400$        &$400$       &$400$       \\
$x_h$&$0.0751~(6)$&$0.0857~(6)$&$0.0990~(6)$&$0.1108~(6)$ &$0.1290~(8)$    \\
\hline\hline
\end{tabular}
\end{center}
\end{table}

Next, we fitted the $g_p (r=L/2)$ data at $J=3.0$ by
\begin{equation}
g_p=L^{-2x_h}(g_0+g_1 L^{-1}+g_2 L^{-2})\; ,
\label{fit_gp}
\end{equation}
where $x_h$ is the associated scaling dimension. The terms with $g_1$
and $g_2$ describe the finite-size corrections, and the correction
exponents are simply set at $-1$ and $-2$, respectively.
The results are shown in Table~\ref{tab_xJ3}.

We conjecture that the fractal dimension $2-x_h$ of the percolation
clusters at $K_{\rm c} (J)$ assumes the exact BKT value with $x_h=1/8$. 
This conjecture is based on the BKT-like behavior of the percolation
transition in the low-temperature range, and on the numerical evidence
for $x_h$ obtained from the correlation $g_p (r=L/2)$ in Table~\ref{tab_xJ3}.
First, the percolation in the low-temperature range seems to be BKT-like.
Second, the fit results for the scaling dimension $x_h$,
when interpolated to $K_{\rm c}$ as given in Table \ref{tab_Kcsq},
yield a value close to $1/8$. The data for the scaled quantity
$g_p (r=L/2) L^{1/4} (\ln L)^{-1/8}$, shown in Fig.~\ref{fig_gpsq2}
for $J = 3.0$, confirm the existence of intersections, apparently
converging to the same value of $K$ as those in Fig.~\ref{fig_qpsq}.
Furthermore we found that the data for $g_p(r=L/2)$ in the interval
$0.47 \leq K \leq 0.515$ at $J = 3.0$ are well described by 
Eq.~(\ref{fit_m2}) for finite sizes in the range $32 \leq L \leq 400$.
This fit yields an estimate for the  percolation threshold at
$K_{\rm c} (J=3.0) =0.504~(8)$. 

\begin{table}[htp]
\vspace*{0cm} \hspace*{-0cm}
\begin{center}
\caption{Percolation threshold $K_{\rm c} (J)$ for various values of
the XY coupling strength $J$.}
\label{tab_Kcsq}
\begin{tabular}{c|c|cccccc}
\hline\hline
square &$J$   & $3.0$    &$2.4$       &$1.8$       &$1.4$       &$1.2$  &$1.12$     \\
 &$K_{\rm c}$&$0.504~(8)$&$0.553~(5)$&$0.646~(9)$&$0.785~(9)$&$0.946~(8)$ &$1.120~(9)$\\
\hline
triangular &$J$& $2.2$   &$1.8$        &$1.4$        &$1.0$      &$0.8$   &$0.6824$    \\
&$K_{\rm c}$&$0.300~(3)$&$0.310~(1)$&$0.356~(4)$&$0.424~(9)$&$0.520~(9)$ &$0.675~(9)$ \\
\hline\hline
\end{tabular}
\end{center}
\end{table}

We also performed simulations at $J=2.4$, $1.8$, $1.4$, and $1.2$,
and observe a behavior similar as that described above for $J=3.0$.
On the basis of a fit of the $g_p L^{1/4} (\ln L)^{-1/8}$ data by
Eq.~(\ref{fit_m2}), we obtain the associated percolation thresholds,
which are shown in Table~\ref{tab_Kcsq}. 
Next, we fitted Eq.~(\ref{fit_gp}) to the $g_p (r=L/2)$ data at various 
points  $(J,K)$ for $J \geq J_{\rm c}$ and $K \geq K_{\rm c} (J)$. 
The results are shown in Table~\ref{tab_xJK}.

In addition, we carried out simulations at $J=1.12$, very close to the 
thermal critical point $J_c=1.124~(3)$. The ratio $Q_p (K)$ appears to
behave similarly as in Fig.~\ref{fig_qpsq}, which suggests a BKT-like
percolation transition. The estimated threshold $K_c=1.120~(9)$ agrees with
the critical point $J_c$. This fits well with the continuation of the 
$K_c(J)$ line in Fig.~\ref{fig_phase}. Further, the numerical result
for the fractal dimension of the percolation clusters at $K = J_c$ is
consistent with the BKT value $2-x_h=15/8$. 

The results in Table~\ref{tab_Kcsq} and~\ref{tab_xJK} are summarized 
in Figs.~\ref{fig_phase} and~\ref{fig_xh}, respectively.

\begin{table}[t]
\vspace*{0cm} \hspace*{-0cm}
\begin{center}
\caption{Results for the scaling dimension $x_h$ at various points 
$(J,K)$ for the square lattice.} 
\label{tab_xJK}
\begin{tabular}{c|c|ccccc}
\hline\hline
$J=3.0$ &$K$ & $3.0$ &$1.8$       &$1.2$       &$0.9$       &$0.72$       \\
&$x_h$ &$0.02916~(2)$&$0.0345~(6)$&$0.0412~(6)$&$0.0492~(6)$&$0.0590~(6)$ \\
\cline{2-7}
        &$K$ &$0.60$ &$0.555$     &$0.525$     &$0.510$     &$0.495$      \\
&$x_h$ &$0.0751~(6)$ &$0.0857~(6)$&$0.0990~(6)$&$0.1108~(6)$&$0.1290~(8)$ \\
\hline\hline 
$J=2.4$ &$K$ & $2.4$ &$1.44$      &$1.2$       &$0.96$      &$0.84$       \\
&$x_h$ &$0.03748~(5)$&$0.0460~(6)$&$0.0491~(6)$&$0.0560~(6)$&$0.0628~(6)$ \\
\cline{2-7}
        &$K$ & $0.72$&$0.648$     &$0.60$      &$0.576$     &$0.552$      \\
&$x_h$  &$0.0719~(6)$&$0.0821~(6)$&$0.0925~(6)$&$0.1010~(6)$&$0.1151~(6)$ \\
\hline\hline  
$J=1.8$ &$K$ & $1.8$ &$1.26$      &$1.08$      &$0.9 $      &$0.81$       \\
&$x_h$ &$0.05282~(5)$&$0.0614~(6)$&$0.0672~(6)$&$0.0759~(6)$&$0.0832~(6)$ \\
\cline{2-7}
        &$K$ & $0.72$&$0.684$     &$0.666$     &$0.648$     &             \\
 &$x_h$ &$0.0954~(6)$&$0.1028(6)$ &$0.1069~(6)$&$0.1140~(6)$&$         $  \\
\hline\hline
$J=1.4$ &$K$ & $1.4$ &$1.19$      &$0.98$      &$0.854$     &$0.798$      \\
&$x_h$ &$0.07386~(5)$&$0.0805~(6)$&$0.0915~(6)$&$0.1033~(6)$&$0.1125~(6)$ \\
\cline{2-7}
        &$K$ &$0.784$&$0.77 $     &$0.756$     &$     $     &             \\
 &$x_h$ &$0.1157~(6)$&$0.1202~(6)$&$0.1250~(6)$&$        $  &$         $  \\
\hline\hline
$J=1.2$ &$K$ & $1.8$ &$1.44$      &$1.2 $      &$1.14 $     &$1.08 $      \\
&$x_h$&$0.0798~(6)$ &$0.0873~(6)$&$0.09610~(6)$&$0.0988~(6)$&$0.1028~(6)$ \\
\cline{2-7}
        &$K$ &$1.02$&$0.96 $      &$     $     &$     $     &             \\
&$x_h$&$0.1084~(6)$  &$0.1159~(6)$&$         $ &$         $ &$         $  \\
\hline\hline
$J=1.12$ &$K$ & $3.0$ &$2.5$      &$2.0 $      &$1.5 $     &$1.21 $      \\
&$x_h$&$0.0784~(3)$ &$0.0821~(3)$&$0.0876~(4)$&$0.0984~(5)$&$0.1108~(3)$ \\
\cline{2-7}
        &$K$ &$1.18$&$1.15 $      &$     $     &$     $     &             \\
&$x_h$&$0.1132~(3)$  &$0.1162~(5)$&$         $ &$         $ &$         $  \\
\hline\hline
\end{tabular}
\end{center}
\end{table}

\subsection{Percolation on the triangular lattice}\label{trik=1}

\subsubsection{Matching property}
The matching property~\cite{Sykes-64,Essam-87} plays an important role
in the determination of the site percolation thresholds of several
two-dimensional lattices; here we briefly review this subject.
For a given planar lattice 
${\mathcal P}\equiv ({\mathcal V}, {\mathcal B})$, where ${\mathcal V}$ 
is the set of lattice sites and ${\mathcal B}$ is the edge set, one does
the following:  1), select parts of the faces of $\mathcal P$, and fill in 
all the ``diagonals'' in those faces. This yields lattice ${\mathcal L}
\equiv ({\mathcal V}, {\mathcal B}+{\mathcal A})$, where ${\mathcal A}$ 
represents the set of all added diagonal edges. 2), select the faces
that are not picked up in step 1), and fill in all the diagonals in 
these faces. One has lattice then ${\mathcal L}^* \equiv 
({\mathcal V},{\mathcal B}+{\mathcal A} ^*) $ 
with ${\mathcal A}^*$ the set of diagonals drawn in step 2). One 
calls lattices ${\mathcal L}$ and ${\mathcal L}^*$ are matching to 
each other; note that ${\mathcal L}$ and ${\mathcal L}^*$ may be
non-planar. Since no ``diagonal'' can be filled in a triangle, 
the triangular lattice is self-matching. It can be shown that, 
for the site-percolation problem, the cluster numbers per 
site $\kappa (p)$ and $\kappa^* (1-p)$ on a pair
of matching lattices ${\mathcal L}$ and ${\mathcal L}^*$ satisfy
\begin{equation}
\kappa (p) - \kappa^* (1-p) = \phi (p) \; ,
\label{mat}
\end{equation}
where $p$ is the site-occupation probability and $\phi (p)$ is a finite
polynomial (it is termed ``matching polynomial''). Equation~(\ref{mat})
indicates that, if the cluster-number density $\kappa$ on the lattice
${\mathcal L}$ exhibits a singularity at a site occupation probability
 $p$, the same singularity will also occur 
in $\kappa^*$ on ${\mathcal L}^*$ at $1-p$.  Together with the 
plausible assumption that 
there is only one transition, the matching argument yields that 
the percolation threshold is $p_{\rm c}=1/2$ for all self-matching lattices
like the triangular lattice; further, it requires that $\phi (p=1/2)=0$,
which is indeed satisfied by the result~\cite{Essam-87}
$\phi (p)=p (1-p) (1-2p)$ for self-matching lattices.
An important feature of the matching argument is that it is still valid
in the presence of interactions, as long as these interactions are
symmetrical under the interchange of occupied and unoccupied sites.

\subsubsection{Percolation at $\tanh K=1$}

As mentioned in Sec.~\ref{persq_htr}, the case $\tanh K = 1$, $J = 0$
in the present percolation process corresponds with the case
$p = 1/2$ for the standard-site percolation.
The standard site-percolation threshold for the triangular 
lattice is $p_{\rm c} = 1/2$, thus the percolation threshold of the 
present percolation problem at $J = 0$ is $\tanh K =1$.

Since the matching argument is independent of the coupling $J$, and
no spontaneous symmetry breaking occurs in the two-dimensional XY model,
$\tanh K =1$ describes a critical line for finite $J$. 
Further, we expect that, in the high-temperature 
range $J < J_{\rm c}$, the percolation transition is in the universality
class of standard uncorrelated percolation, since there is no long-range
spin-spin correlation. 

\begin{figure}[t]
\vspace*{0cm} \hspace*{-0cm}
\begin{center}
\includegraphics[width=8.0cm]{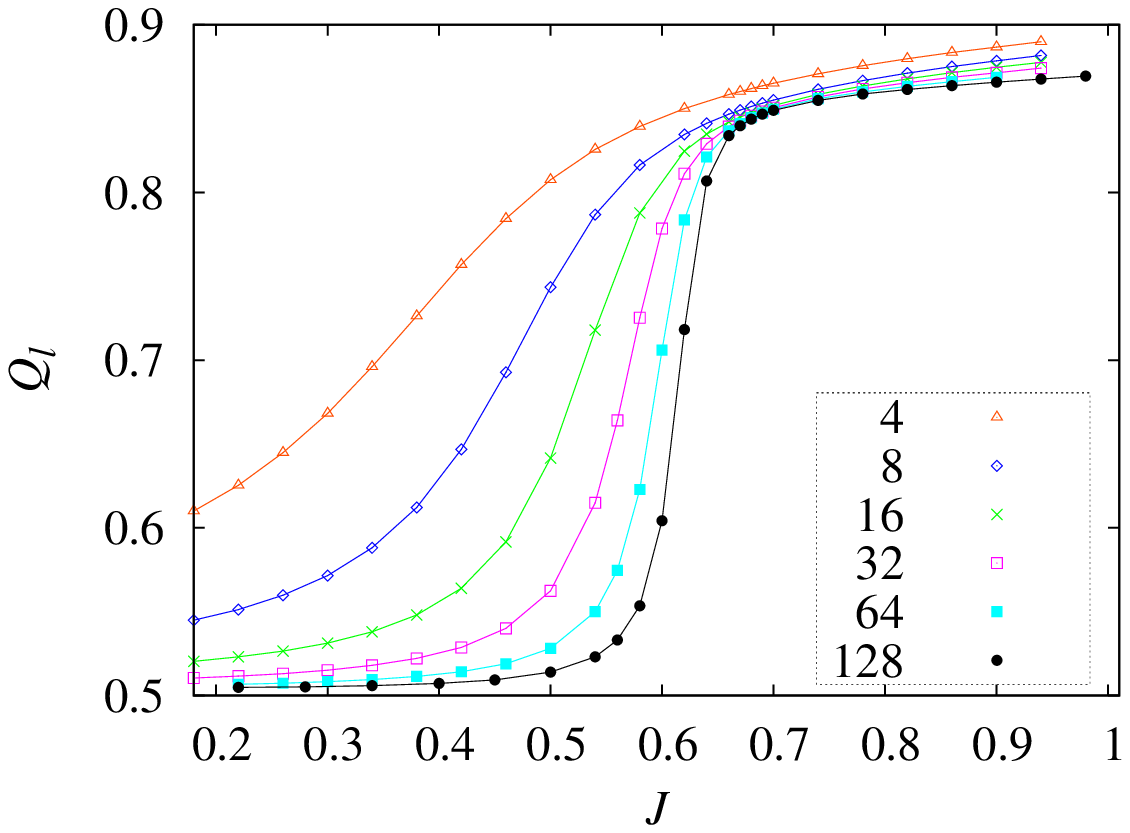}

\caption{(color online).
Ratio $Q_l$ vs. coupling strength $J$ for the triangular lattice at
$\tanh K=1$. The lines connecting the data points are added for clarity.}
\label{fig_qltr}

\includegraphics[width=8.0cm]{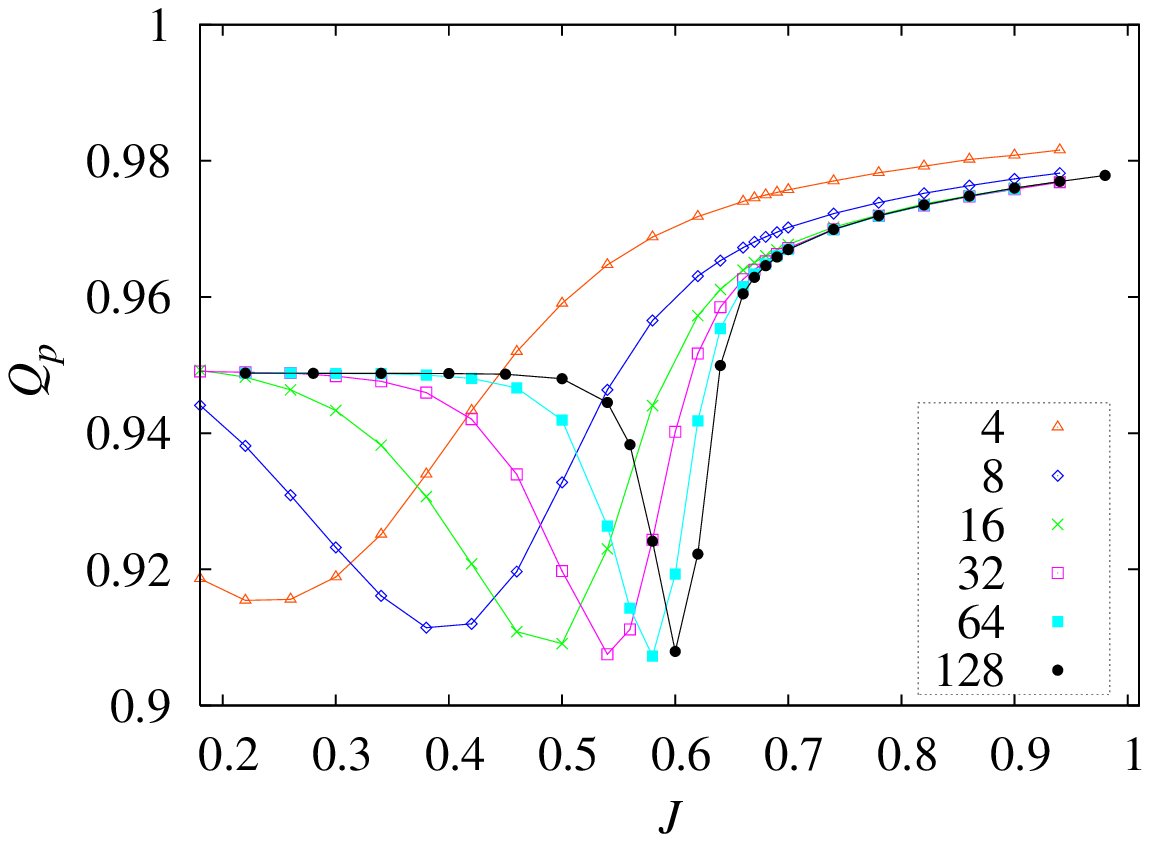}

\caption{(color online).
Ratio $Q_p$ vs. coupling strength $J$
 on  the triangular lattice at $\tanh K=1$.
The lines connecting the data points are added for clarity.}
\label{fig_qptr}

\includegraphics[width=8.0cm]{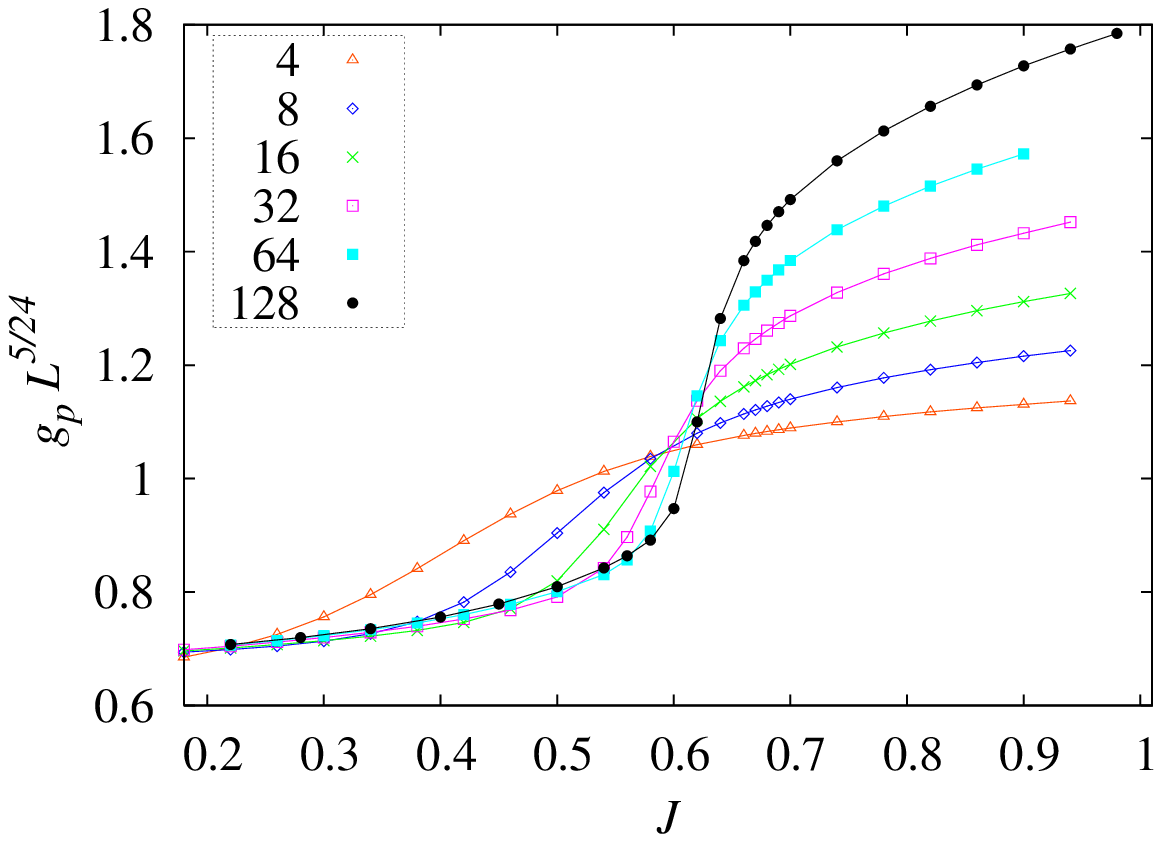}
\caption{(color online).
Scaled correlation $g_p (L/2) L^{2x_h}$ 
vs. coupling strength $J$ for the triangular lattice at $\tanh K=1$,
with $x_h=5/48$ which applies to the uncorrelated percolation universality
class.  The lines connecting the data points are added for clarity.}
\label{fig_gptr}
\end{center}
\end{figure}

Figure~\ref{fig_qltr} shows the data for the ratio $Q_l$, 
which is defined by Eq.~(\ref{def_Ql}) on the basis of the size
distributions of the percolation clusters.  For $J > J_{\rm c}$, $Q_l$
approaches a $J$-dependent value which is clearly smaller than 1.
This implies the absence of an infinite cluster
that occupies a finite fraction of the whole lattice. The singularity 
at the thermal transition point $J_{\rm c}$ is reflected by the jump
that develops near $J_{\rm c} \approx 0.68$.

Figure~\ref{fig_qptr} shows the data for the ratio $Q_p$.
For $J<J_{\rm c}$, $Q_p$ converges to a universal value $Q_{pc}\approx 0.95$
(note that this value differs from $Q_{pc}=0.872776~(3)$~\cite{Zhang-08}
for standard percolation on the triangular lattice, due to the
difference mentioned in the first paragraph of Sec.~\ref{persq_htr}). 
For $J > J_{\rm c}$, $Q_p$ approaches a $J$-dependent value smaller than 1; 
we thus expect that the correlation $g_p (r)$ decays algebraically rather
than exponentially. 
The thermal transition at $J_{\rm c}$ is reflected by the rapid variation
of $Q_p$ near $J_{\rm c} \approx 0.68$ for large $L$.

The data for the scaled correlation $g_p (L/2)$ at $\tanh K=1$ 
are shown in Fig.~\ref{fig_gptr} as a function of $J$, with 
$x_h = 5/48$ for the uncorrelated percolation universality.
The convergent behavior for $J < J_{\rm c}$ as a function of $L$ confirms
that the transition in this range belongs to the standard percolation
universality class.
The intersections roughly represent the thermal transition point $J_{\rm c}$.

\subsubsection{Percolation at $\tanh K \neq 1$}

\begin{figure}[t]
\vspace*{0cm} \hspace*{-0cm}
\begin{center}
\includegraphics[width=7.5cm]{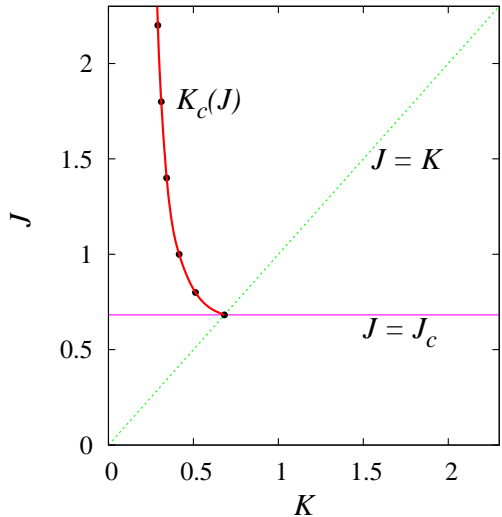}
\caption{(color online).
Phase diagram in the $J-K$ parameter space for the
XY model on the triangular lattice. The horizontal 
line is the thermal BKT transition, and the diagonal line is for $K=J$,
where the percolation clusters are just those formed by the Monte Carlo
cluster algorithm. 
In addition there is a percolation line for $J < J_{\rm c}$ at
$\tanh K =1$, outside the range of this figure.
The line connecting the data points is added for clarity.}
\label{fig_phase_tri}
\end{center}
\end{figure}

Following similar procedures as in Sec.~\ref{persq}, we obtain a
percolation line $K_{\rm c} (J)$ in low the temperature range
$J > J_{\rm c}$ for the triangular lattice. For $J < J_{\rm c}$ we do
not find a percolation threshold at finite values of $K$.
The numerical results are shown in
Table~\ref{tab_Kcsq} and Fig.~\ref{fig_phase_tri}.  It is observed that
the percolation in the range $J > J_{\rm c}$ is also BKT-like, with a
fractal dimension $2 - x_h = 15/8$  at $K_{\rm c} (J)$, and a scaling
dimension $x_h$ depending on parameters $K$ and $J$ in the range
$K > K_{\rm c} (J)$.  This is consistent with the results for the square
lattice in Sec.~\ref{persq}.

\section{Discussion}\label{sec_dis}

Since spins in the same cluster formed during the simulations must
have $x$-components of the same sign, the absence of a spontaneous
magnetization~\cite{Mermin-66} in the XY model means that the density
of the largest cluster in the thermodynamic limit is also restricted
to be zero, at least for finite values of $J$. The same restriction
thus applies to percolation clusters formed with $K=J$ in Eq.~(\ref{pij}),
and it must also hold for $K<J$. 
The absence of a nonzero density of the largest percolation cluster
is in agreement with the interpretation of the percolation
transitions for $K<J$ described in Sec.~\ref{sec_per} as BKT-like.

The results presented in Sec.~\ref{persq} for the square lattice indicate
that, in addition to the line $K_c(J)$ with $J\geq K$, also the line
$J=J_c$, $K>J_c$ represents a percolation threshold.  As can be seen
from the data points for $J=1.12 \approx J_c$ in Fig.~\ref{fig_xh},
the magnetic exponent depends on $K$ along the latter line, which is thus
a ``nonuniversal'' line of percolation transitions, and the bond dilution
field parametrized by $K$ is truly marginal. The existence of a BKT 
transition induced by varying $K$ at the point $K=J=J_c$ corresponds
with a marginally relevant bond-dilution field in the $K<J_c$ direction.

As mentioned earlier, for the triangular lattice with $J < J_{\rm c}$, the
line $\tanh K = 1$ is critical, and belongs to the standard percolation
universality class. 
The continuation of the $\tanh K = 1$ line to $J \ge J_{\rm c}$ is also
critical (in the sense that the correlation functions display algebraic
decay), but with a $J$-dependent critical exponent. 

In order to obtain some more information on the dependence of the present
percolation problem on the coordination number $z$, we also simulated the
$z=18$ equivalent-neighbor XY model on the triangular lattice, which has 
equal nearest-, second-nearest- and third-nearest-neighbor interactions.
The procedure outlined in Sec.~(\ref{sec_cri}) yielded an estimate of
the thermal transition at $J_{\rm c} = 0.162~(2)$.  
As expected, a Monte Carlo analysis of the percolation problem with $z=18$ 
showed the existence of a critical line $K_{\rm c}(J)$ in the high-temperature
phase $J < J_{\rm c}$, belonging to the standard percolation universality
class. When $J$ approaches $J_{\rm c}$, the $K_c(J)$ line bends
toward large values of $K$. This suggests that $K_c(J)$ line ends at
$K \rightarrow \infty$ for $J=J_c$.
For $K>K_{\rm c}(J)$ there is clear evidence for the existence of a
percolation cluster with a finite density in the limit of large $L$.

In the low-temperature range $J > J_{\rm c}$, we found, just as for the
models with nearest-neighbor interactions, a BKT-like transition line
$K_{\rm c}(J)$, as in Figs.~\ref{fig_phase} and \ref{fig_phase_tri}.
In spite of the relatively large coordination number $z=18$, no evidence is 
found for a percolation cluster of a nonzero density, even at $\tanh K = 1$.
Although the spin-spin correlations in the algebraic XY phase for
$J>J_{\rm c}$ stimulate the percolation transition in the sense that
it occurs at smaller values of $K$ when $J$ increases, it also appears
that they obstruct the formation of a percolation cluster with a
non-zero density.

Besides the rule based on Eq.~(\ref{pij}), other procedures for placing
bonds may be applied. 
For instance, as mentioned in the Introduction, Wang et al.~\cite{Wang-10}
placed percolation bonds between neighboring XY spins if their orientations
differ less than a given threshold, and found percolation transitions
in the uncorrelated percolation universality class for all XY couplings.
Another possibility is to place percolation bonds with probabilities
given, instead of Eq.~(\ref{pij}), by
$p_{ij}=\max (0, 1-e^{-2K \vec{s}_i \cdot \vec{s}_j})$.
In that case, we expect percolation transitions similar to those of
Ref.~\onlinecite{Wang-10}, 
including transitions in the low-temperature range $J > J_{\rm c}$
of the XY model. Indeed, a preliminary Monte Carlo
analysis of this problem \cite{HD} confirms the existence of such
transitions in the universality class of uncorrelated percolation.

Finally we remark that recently a percolation problem was formulated
on the basis of the O($n$) loop configurations~\cite{Ding-09}. 
Since the XY model is equivalent with the O(2) model, another way thus
arises to introduce percolation in XY-type models. Although this seems
a very different approach, it reproduces our result that a marginally
relevant dilution field exists at the BKT transition.

\acknowledgments
We are indebted to Prof. B. Nienhuis for valuable discussions.
This work was supported in part by the
National Nature Science Foundation of China under Grant No.\;10975127, the
Anhui Provincial Natural Science Foundation under Grant No.\;090416224,
and the Chinese Academy of Sciences.

\end{document}